\documentclass[useAMS,usenatbib,usegraphicx]{mn2e}

\title[$z=6.5$ Ly$\alpha$ emitters in the SSA22 field]{Ly$\alpha$
emitters at $z=6.5$ in the SSA22 field: An area more neutral or void at
the end of the reionization epoch\thanks{This work is
based on data collected at Subaru Telescope and obtained from the SMOKA,
which is operated by the Astronomy Data Center, National Astronomical
Observatory of Japan.}}
\author[E.\ Nakamura et al.]{\parbox[t]{\textwidth}{\vspace{-1cm}
E. Nakamura$^{1}$, 
A. K. Inoue$^{2}$\thanks{E-mail: akinoue@las.osaka-sandai.ac.jp},
T. Hayashino$^{1}$, M. Horie$^{1}$, K. Kousai$^{1}$, T. Fujii$^{1}$, 
and Y. Matsuda$^3$}\\
$^{1}$Research Center for Neutrino Science, Graduate School of Science,
Tohoku University, Aramaki, Aoba-ku, Sendai 980-8578, Japan\\
$^{2}$College of General Education, Osaka Sangyo University, 
3-1-1, Nakagaito, Daito, Osaka 574-8530, Japan\\
$^{3}$Department of Physics, Durham University, South Road, Durham DH1
3LE}
\begin{document}

\date{}

\pagerange{\pageref{firstpage}--\pageref{lastpage}} \pubyear{2009}

\maketitle

\label{firstpage}

\begin{abstract}
 We present results of a survey of Lyman $\alpha$ emitters (LAEs) at
 $z=6.5$ which is thought to be the final epoch of the cosmic
 reionization. In a $\approx530$ arcmin$^2$ deep image of the SSA22
 field taken through a narrowband filter NB912 installed in the
 Subaru/Suprime-Cam, we have found only 14 LAE candidates with 
 $L_{\rm Ly\alpha}\ga3\times10^{42}$ erg s$^{-1}$. Even applying the
 same colour selection criteria, the number density of the LAE
 candidates is a factor of 3 smaller than that found at the same
 redshift in the Subaru Deep field (SDF). Assuming the number density in
 the SDF is a cosmic average, the probability to have a number density
 equal to or smaller than that found in the SSA22 field is only 7\% if
 we consider fluctuation by the large-scale structure (i.e. cosmic
 variance) and Poisson error. Therefore, the SSA22 field may be a rare
 void at $z=6.5$. On the other hand, we have found that the number
 density of $i'$-drop galaxies with $25.5<z'<26.0$ in the SSA22 field
 agrees well with that in the SDF. If we consider a scenario that a
 larger neutral fraction of intergalactic hydrogen, $x_{\rm HI}$, in the
 SSA22 field obscures a part of Ly$\alpha$ emission, $x_{\rm HI}$ in the
 SSA22 field should be about 2 times larger than that in the SDF. This
 can be translated into $x_{\rm HI}<0.9$ at $z=6.5$ in the SSA22
 field. A much larger survey area than previous ones is required to
 overcome a large fluctuation reported here and to obtain a robust
 constraint on $x_{\rm HI}$ at the end of the reionization from LAEs.
\end{abstract}


\section{Introduction}

Survey of Ly$\alpha$ emitters (LAEs) at $z\ga6$ has been significantly
developing in the last decade since the initial discovery papers 
\citep{hu99,rho01,kod03,iye06}. At $z\simeq5.7$, many LAE surveys with
the Subaru/Suprime-Cam (S-Cam; \citealt{miy02}) have been made so far 
\citep{hu04,aji03,aji06,shi06,mur07,ouc08}. In particular, \cite{shi06}
have performed a deep LAE survey at $z=5.7$ in the Subaru Deep Field
(SDF) and present a seminal luminosity function (LF) of Ly$\alpha$
emission at the redshift. \cite{mur07} have performed a very wide
($\approx2$ deg$^2$) survey in the COSMOS field and found no evidence of
a large-scale clustering of bright ($L_{\rm Ly\alpha}\ga1\times10^{43}$
erg s$^{-1}$) LAEs at $z=5.7$. Finally, \cite{ouc08} have reported 
results of a wide ($\approx1$ deg$^2$) and deep survey in the
Subaru/XMM-Newton Deep Survey (SXDS) field and confirmed 
the LF by \cite{shi06}. In addition, \cite{ouc08} report a factor of 5
variation of the number density of LAEs among their 5 fields of view of
S-Cam.

On the other hand, few surveys had been reported to date at
$z\simeq6.5$ \citep{tan05,hu05,hu06} and a significant improvement is
arising \citep{ouc10,hu10}. A largest sample of LAEs at $z=6.5$ reported
so far is the one in the SDF presented by \cite{tan05}.
Based on this sample, \cite{kas06} have constructed the Ly$\alpha$ LF at
the redshift and found that a significant reduction at the bright-end
relative to the $z=5.7$ LF by \cite{shi06}. They interpret this
reduction as a signature of the end epoch of the cosmic reionization.

The cosmic reionization is one of the most drastic event in the history
of the universe and it happened at $z\sim10$ \citep{dun09}. The
reionization was probably triggered by the first-generation of stars and
galaxies. Thus, the event contains the information of the structure 
formation in the early universe. The subsequent galaxy formation
proceeded in the reionized universe and was significantly affected by
the ionizing background radiation. This is a complex astrophysical
process (or feedback) which is one of the most important issues to be
resolved in observational cosmology \citep[e.g.,][]{loe01}.

The discovery of the Gunn-Peterson trough \citep{gun65} in spectra of
QSOs and a GRB at $z\sim6$ \citep{bec01,tot06} marks the end of the
reionization epoch \citep[e.g.,][]{fan06}. Furthermore, \cite{kas06}
argue that the evolution of Ly$\alpha$ LF between $z=5.7$ and $z=6.5$
which they found is another signature of the end epoch; since the
measurements of ultraviolet (UV) LF of the LAEs agree very well between
the two redshifts, the apparent faintness of Ly$\alpha$ emission of
$z=6.5$ LAEs is caused by an increase of attenuation by neutral hydrogen
remained in the intergalactic medium (IGM) at $z>6$.

A significant variation in the IGM transmission found in the spectra of
$z\sim6$ QSOs suggests inhomogeneous reionization \citep{djo06}. 
Numerical simulations also show inhomogeneity of the reionization
process \citep[e.g.,][]{cia03}. This means that there may be a large
variation in the end epoch of the reionization among lines of sight. If
it is true, the number density of $z=6.5$ LAEs may variate significantly
among different survey fields. Indeed, \cite{hu06} report such a large
variation between two fields of view of S-Cam. Therefore, another survey
of $z=6.5$ LAEs similar to \cite{tan05} is indispensable to discuss such
a field-to-field variation.

This paper presents a photometric sample of $z=6.5$ LAEs found in the
archival imaging data taken with S-Cam towards the SSA22 field. Some of
the LAEs have been confirmed by spectroscopy in \cite{hu10}. Then, we
compare the number density of the LAEs in the SSA22 field with that in
the SDF in order to discuss the field-to-field variation
quantitatively. The structure of the rest of this paper
is as follows; first, we describe the imaging data used in this paper in
\S2. Then, we present the photometric sample of $z=6.5$ LAEs in the
SSA22 in \S3. In \S4, we compare the Ly$\alpha$ LF derived in the SSA22
with that in the SDF, and also compare the number density of $i'$-drop
galaxies in the two fields. In \S5, we discuss the cause of the small
number of LAEs in the SSA22. The final section provides the conclusions
of this paper.

The cosmology adopted in this paper is a flat $\Lambda$CDM cosmology with 
$h=0.7$, $\Omega_{\rm M}=0.3$, and $\Omega_\Lambda=0.7$. All magnitudes
are described by the AB system: ${\rm AB}=-2.5\log f_\nu-48.60$, where 
$f_\nu$ in unit of erg s$^{-1}$ cm$^{-2}$ Hz$^{-1}$ \citep{oke74}.

\section{Imaging data}

We have collected imaging data of the SSA22 field 
($\alpha=22^h17^m34^s$, $\delta=+00^\circ17'00''$; J2000) taken with
Subaru/Suprime-Cam (S-cam; \citealt{miy02}) from the archive data
system, SMOKA (Subaru-Mitaka-Okayama-Kiso Archive System;
\citealt{bab02}). The collected broadband data are $B$, $V$, $R$,
$i'$, and $z'$. In order to select $z\simeq6.5$ LAEs, we use the imaging
data through a narrowband NB912 (central wavelength of 9,139 \AA\ and
full width at half maximum [FWHM] of 134 \AA) whose efficiency curve is
shown in Figure 1. Table~1 is a summary of the imaging data.

\begin{figure}
 \begin{center}
  \includegraphics[width=6cm]{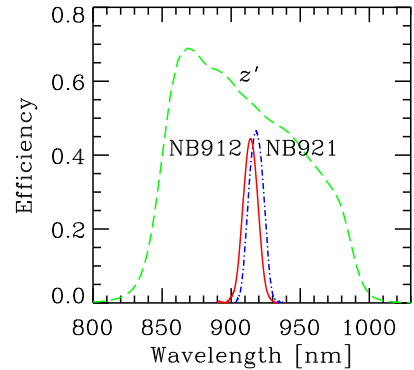}
 \end{center}
 \caption{Efficiency curves of the narrowband filters to select
 $z\simeq6.5$ LAEs. NB912 was used for the SSA22 field (this work) but
 NB921 was used for the SDF (Taniguchi et al.~2005). The efficiencies
 include the CCD quantum efficiency as well as the filter
 transparencies.}
\end{figure}

\begin{table*}
 \caption[]{Imaging data of the SSA22 field used in this paper.}
 \setlength{\tabcolsep}{3pt}
 \footnotesize
 \begin{minipage}{\linewidth}
  \begin{tabular}{lcccc}
   \hline
   Filter & Observed year (P.I.) & Exposure time (h) & PSF FWHM ($''$) 
   & 1-$\sigma$ limit (AB)$^a$\\
   \hline
   $B$ & 2008 (Y.~Nakamura) & 2.6 & 0.78 & 29.08 \\
   $V$ & 2002 (T.~Hayashino), 2003 (P.~Capak), 2008 (Y.~Nakamura) 
       & 2.1 & 0.82 & 28.85 \\
   $R$ & 2000, 2001 (E.~Hu) & 2.9 & 1.06 & 28.22 \\
   $i'$ & 2001 (P.~Capak) & 1.4 & 0.78 & 28.28 \\
   $z'$ & 2001 (Y.~Komiyama), 2002, 2003 (P.~Capak) & 2.9 & 0.77 & 27.77 \\
   NB912 & 2002, 2003 (P.~Capak) & 12.7 & 0.77 & 27.79 \\   
   \hline
  \end{tabular}

  $^a$ Aperture diameter is twice of the PSF FWHM. The Galactic dust
  extinction is corrected based on Schlegel et al.~(1998).
 \end{minipage}
\end{table*}%

All the data were reduced using {\sc sdfred} \citep{yag02,ouc04} and
{\sc iraf} with the standard manner. In this paper, we selected only
frames taken under a good seeing condition. Then, FWHMs of stellar
objects in the final images are relatively good as summarised in Table~1,
except for $R$ band.  The relative accuracy in astrometry in the
final images is less than 0.5 pix (pixel scale is $0.''202$). The flux
density calibrations of $z'$ and NB912 images were done by a photometric
standard star GD 50 and a spectroscopic standard star G 93-48,
respectively. The calibrations of other broadband images were done so as
that the colour distributions of $z'$ around 25.5 AB objects in the
SSA22 agree with those in the SDF reported by \cite{kas04}, after the
correction of the Galactic dust extinction based on \cite{sch98}.

The data of $z'$ and NB912 were taken with two position angles with a
difference of 90$^\circ$. While the S-cam has a field of view of
$34'\times27'$, therefore, the area with a good homogeneous quality in
the final co-added image is reduced to $22.'7\times23.'6$ for the two
filters. We use only this area to search for LAEs at $z\simeq6.5$. The
limiting magnitudes of $z'$ and NB912 in Table~1 are also measured in
this area. The co-moving depth of our survey is 41.6 $h_{0.7}^{-1}$ Mpc
since NB912 effectively captures Ly$\alpha$ emission line between
$z=6.46$ and $z=6.57$. Finally, our survey volume is $1.3\times10^5$
$h_{0.7}^{-3}$ Mpc$^3$ in co-moving unit.

As a comparison field, we adopt the SDF 
($\alpha=13^h24^m39^s$, $\delta=+27^\circ29'25''$; J2000). \cite{kas04}
published the reduced and calibrated images of $B$, $V$, $R$, $i'$,
$z'$, and two narrowband NB816 and NB921. The NB921 has the central
wavelength of 9,196 \AA\ and FWHM of the transmission of 132 \AA, which
are quite similar to those of NB912 as shown in Figure~1. \cite{tan05}
and \cite{kas06} reported their results of LAE survey using NB921 in the
SDF. We use these images and photometric catalogue reported in
\cite{kas04} and in \cite{tan05} for a comparison with our LAE
candidates in the SSA22 field in \S4.

Before selecting LAEs, we should confirm the accuracy of the flux
calibration of our data in the SSA22 field. For this aim, we compare the
$z'$ number count in the SSA22 field with that in the SDF. The result is
shown in Figure~2. We have found an excellent agreement in $22<z'<26$.
The differences in brighter and fainter magnitudes are due to saturation
of counts and detection limits. We have also confirmed an excellent
agreement in NB number counts. This ensures the accuracy of our flux
calibration in the SSA22 field.

\begin{figure}
 \begin{center}
  \includegraphics[width=6cm]{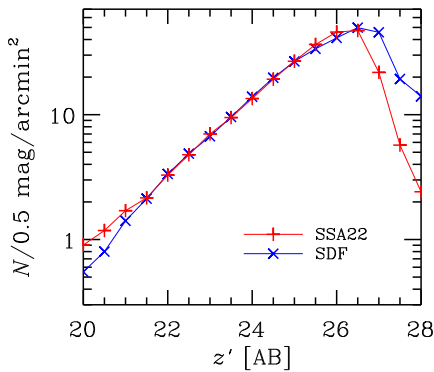}
 \end{center}
 \caption{$z'$ number counts in the SSA22 field (plus) and SDF (cross).}
\end{figure}

\section{Selection of $z=6.5$ LAE candidates}

First, we made object detection in the final NB912 image by
SExtractor \citep{ber96} with the criterion of ``5 connected pixels
above 2-$\sigma$''. We detected 64,787 objects. Then, we selected
objects satisfying all the following criteria: 
\begin{quote}
 \item[(a)] $22.00<$ NB912 $<26.04$ (5$\sigma$), 
 \item[(b)] $z'-{\rm NB912} > 1.0$, 
 \item[(c)] $z'-{\rm NB912} > -2.5\log_{10}(1-3\sigma[F_{z'}/F_{\rm NB912}])$, 
 \item[(d)] $i'-z' > 1.3$ or $i' > 27.53$ (2$\sigma$), 
 \item[(e)] $B>27.89$, $V>27.66$, and $R>27.03$ (3$\sigma$), 
\end{quote}
where $\sigma$ indicates 1-$\sigma$ uncertainty of photometry in each
band and $\sigma[F_{z'}/F_{\rm NB912}]$ means 1-$\sigma$ uncertainty of
the flux density ratio of $z'$ and NB912. The criterion (c) ensures that
the colour excess deviates from a flat $F_\nu$ spectrum by more than 
3-$\sigma$ of the flux density ratio. These criteria are equivalent
with those in the SDF applied by \cite{tan05}, but our $B$, $V$, $R$,
and $i'$ depths are 0.2--0.8 AB shallower than those in the SDF. The
resulting number of objects is 12, after visual check to remove objects
evidently affected by spiders of bright foreground stars. We show the
colour-magnitude diagram, the spatial distribution, and thumbnail images
of the LAE candidates in Figures~3, 4, and 5, respectively. The
coordinate and photometric measurements of these objects are summarised
in Table~2.

\begin{figure}
 \begin{center}
  \includegraphics[width=6cm]{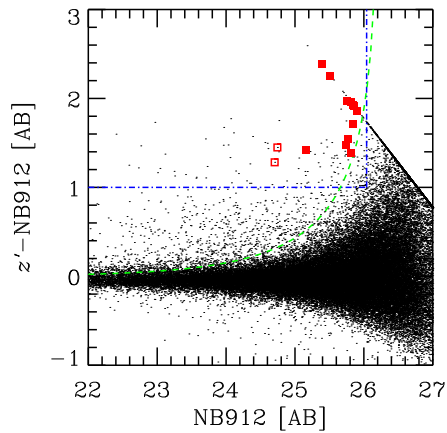}
 \end{center}
 \caption{Colour-magnitude diagram between $z'-{\rm NB912}$ and
 NB912 in the SSA22 field. All objects detected in NB912 are shown by
 small dots. The dot-dashed and dashed lines show the applied colour
 and magnitude cuts and 3-$\sigma$ uncertainty in the flux density ratio
 of $z'$ and NB912, respectively. The selected LAE candidates are shown
 by filled squares. The open squares are one additional LAE candidate
 and one spectroscopic LAE whose photometry is affected by a
 neighbouring object. If objects are not detected in $z'$, the flux
 densities are replaced by 1-$\sigma$ uncertainty in the band.}
\end{figure}

\begin{figure}
 \begin{center}
  \includegraphics[width=6cm]{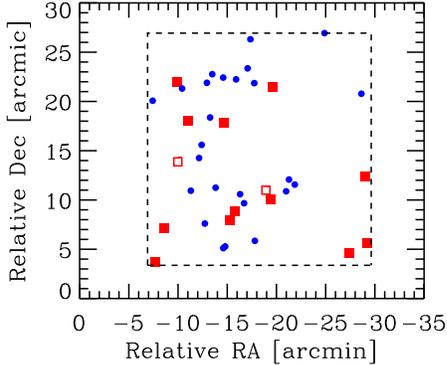}
 \end{center}
 \caption{Spatial distribution of candidates of galaxies at $z\ga6$ in the
 SSA22 field. The filled squares are $z\simeq6.5$ LAE candidates and the
 open squares are one additional candidate and one spectroscopic LAE
 whose photometry is affected by a neighbouring object. The circles are
 $i'$-drop galaxies with $25.5<z'<26.0$. The dashed line show the survey
 area.}
\end{figure}

\begin{figure}
 \begin{center}
  \includegraphics[width=8cm]{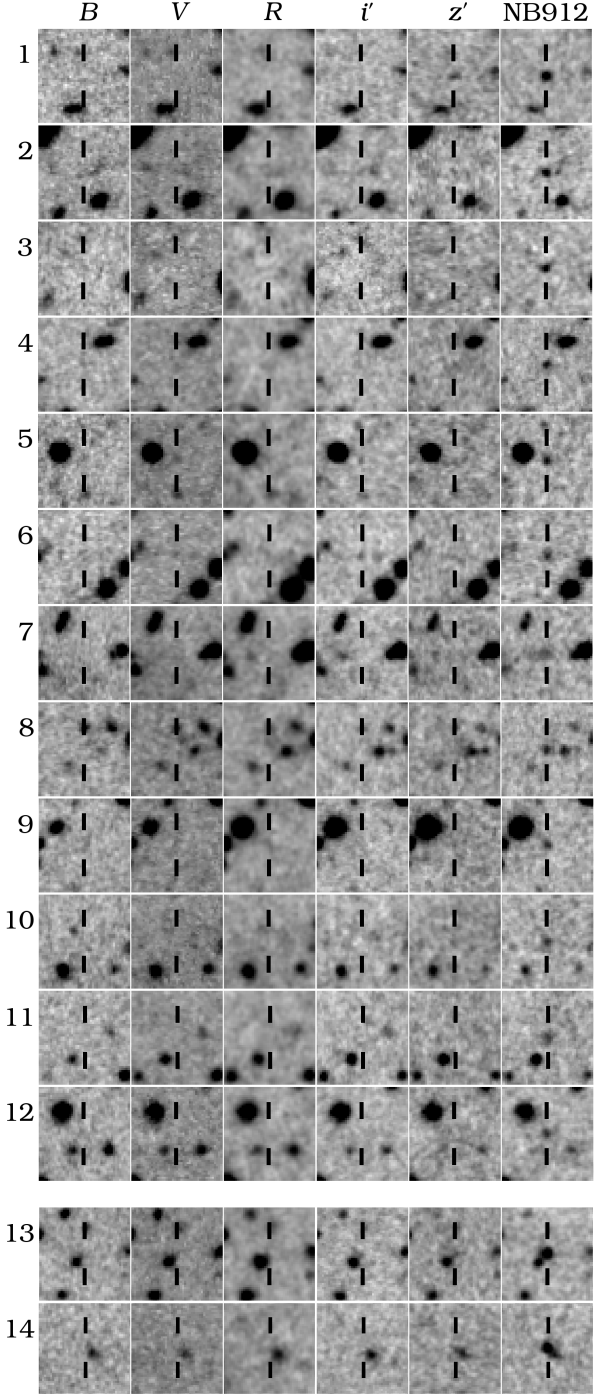}
 \end{center}
 \caption{Thumbnail images of the $z\simeq6.5$ LAE candidates. North is
 up and east is left. The field of view is $9''\times9''$.}
\end{figure}

\begin{table*}
 \caption[]{Observed properties of $z\simeq6.5$ LAE candidates in the
 SSA22 field.}
 \setlength{\tabcolsep}{3pt}
 \footnotesize
 \begin{minipage}{\linewidth}
  \begin{tabular}{lcccccc}
   \hline
   Object & RA & Dec & $i'$ $^a$ & $z'$ $^a$ & NB912 $^a$ & Remarks \\
   & (J2000) & (J2000) & (AB) & (AB) & (AB) & \\
   \hline
   SSA22-z6p5-1 & 22:17:42.84 & +00:18:08.5 
	& $>28.28$ & 26.57 & 25.16 & HC221742-001808 ($z_{\rm sp}=6.4692$)\\
   SSA22-z6p5-2 & 22:16:44.62 & +00:05:54.8 
	& 27.77 & $>27.77$ & 25.39 & \\
   SSA22-z6p5-3 & 22:18:10.79 & +00:04:02.0 
        & $>28.28$ & $>27.77$ & 25.51 & \\
   SSA22-z6p5-4 & 22:17:40.13 & +00:08:14.1 
        & 27.61 & 27.22 & 25.74 & \\
   SSA22-z6p5-5 & 22:16:45.15 & +00:12:38.3 
        & $>28.28$ & 27.72 & 25.76 & \\
   SSA22-z6p5-6 & 22:16:51.72 & +00:04:53.5 
        & 27.60 & 27.32 & 25.77 & \\
   SSA22-z6p5-7 & 22:18:07.05 & +00:07:26.9 
	& $>28.28$ & $>27.77$ & 25.81 & \\
   SSA22-z6p5-8 & 22:17:57.12 & +00:18:19.4 
        & $>28.28$ & 27.21 & 25.81 & \\
   SSA22-z6p5-9 & 22:17:22.73 & +00:21:44.2 
        & 28.23 & $>27.77$ & 25.84 & \\
   SSA22-z6p5-10 & 22:17:23.69 & +00:10:23.0 
        & $>28.28$ & 27.55 & 25.84 & \\
   SSA22-z6p5-11 & 22:17:38.28 & +00:09:09.1 
        & 28.23 & $>27.77$ & 25.85 & HC221738-000909 ($z_{\rm sp}=6.4811$)\\
   SSA22-z6p5-12 & 22:18:01.65 & +00:22:21.1 
        & $>28.28$ & $>27.77$ & 25.91 & HC221801-002220 ($z_{\rm sp}=6.5360$)\\
   \hline
   SSA22-z6p5-13 $^b$ & 22:18:01.41 & +00:14:12.2 
        & 27.15 & 25.99 & 24.71 & \\
   SSA22-z6p5-14 $^b$ & 22:17:25.65 & +00:11:18.4
        & 27.16 & 26.19 & 24.74 & HC221725-001119 ($z_{\rm sp}=6.5142$)\\
   \hline
  \end{tabular}

  $^a$ Magnitudes measured within a circular aperture diameter of twice
  of PSF FWHM of each band. Magnitudes below the 1-$\sigma$ limit are
  denoted by the lower limit.\\
  $^b$ Photometry of this object is affected by a neighbouring object.\\
 \end{minipage}
\end{table*}%

In addition, we found one object which may be a LAE at $z\simeq6.5$ in
a careful visual checking of hundreds objects satisfying the magnitude
and colour cuts (criteria [a]--[c]) but not the drop-out criteria ([d]
and [e]). The photometry in shorter wavelengths of this object is
probably affected by a bright neighbour as found in Figure~5 second line
from the bottom. If the effect by the neighbour was removed, this object
would satisfy all the selection criteria. As shown in Figure~3 by an
open square (brighter one), this object has an excess large enough in
$z'-$NB912. We add this object as No.13 in the final list of our LAE
candidates.

Furthermore, we add another object as No.14 in the LAE list. This object
is reported in \cite{hu10} as a spectroscopic LAE. As shown in Figure~5
bottom line, this object has a close neighbour which is probably low-$z$
object. The separation is very close ($\approx1''$), so that we missed
finding it. The colour excess $z'-$NB912 of this object satisfies with
the criterion (b) shown in Figure~3 (fainter open square).

In our photometric LAE sample, there are 4 objects confirmed to be LAEs
at $z=6.5$ by \cite{hu10} spectroscopically. We note their redshift in
Table~2. Other objects are possibly foreground objects like 
[O {\sc ii}], [O {\sc iii}], or H$\alpha$ emitters. Indeed,
\cite{kas06} have found one [O {\sc iii}] emitter and five unidentified
single emission line objects among 22 LAE candidates which they took the
spectra. The five single emission line objects are either LAE or 
[O {\sc ii}] emitter. The contamination fraction in our LAE candidates
would be similar to that in the SDF sample because the selection
criteria are essentially the same. However, this issue should be
confirmed by a follow-up spectroscopy in future.

\section{Comparison with Subaru Deep Field}

We have found 14 candidates of LAE at $z\simeq6.5$ in the SSA22 field. 
In this section, we compare statistics of our LAE candidates with those
in another large survey in the SDF by \cite{tan05} and \cite{kas06}.

\subsection{Ly$\alpha$ luminosity function}

First, we compare the cumulative LFs of the LAE
candidates in the SSA22 field and in the SDF. While we found only 14
candidates in the SSA22, \cite{tan05} reported 57 candidates in
the SDF\footnote{They published 58 objects but one object, No.22, has
$z'-{\rm NB921}=0.87$ which does not satisfy their criterion
$z'-{\rm NB921}>1.0$. Thus, we remove this object from discussions in
this paper.}. Among their 57 objects, one object was found to be 
a foreground [O {\sc iii}] emitter by follow-up spectroscopy and other
five objects may be also foreground contaminations \citep{tan05,kas06}. 
However, we compare our 14 candidates with their 57 candidates because
we may also have foreground contamination in our 14 photometric samples.

We estimate Ly$\alpha$ flux by the same method adopted in \cite{tan05}: 
\begin{equation}
 f_{\rm Ly\alpha} = f_\lambda^{\rm obs}({\rm NB912})\Delta\lambda_{\rm NB912}
  - f_\lambda^{\rm cont}\Delta\lambda_{\rm NB912}/2\,,
\end{equation}
where $f_\lambda^{\rm obs}({\rm NB912})$ is the observed NB912 flux
density (per \AA), $f_\lambda^{\rm cont}$ is the continuum flux density, 
and $\Delta\lambda_{\rm NB912}=134$ \AA\ is FWHM of the NB912 efficiency
curve. The division by 2 accounts for the continuum break at the
shortward of Ly$\alpha$ emission line because of intergalactic
attenuation. The continuum flux density is estimated from 
\begin{equation}
 f_\lambda^{\rm cont} = \{f_\lambda^{\rm obs}(z')\Delta\lambda_{z'} 
  - f_\lambda^{\rm obs}({\rm NB912})\Delta\lambda_{\rm NB912}\}
  / \Delta\lambda_{z'}^{\rm eff}\,,
\end{equation}
where $f_\lambda^{\rm obs}(z')$ is the observed $z'$ flux density, 
$\Delta\lambda_{z'}=1180$ \AA\ is the band-width of the $z'$ filter, 
and $\Delta\lambda_{z'}^{\rm eff}=450$ \AA\ is an effective band-width
for the continuum, which is the width between the longward edges of $z'$
and NB912 band-widths. Note that we omit a numerical factor multiplied
to $f_\lambda^{\rm obs}({\rm NB912})\Delta\lambda_{\rm NB912}$ in
\cite{tan05} because it is actually the efficiency of the $z'$ filter at
Ly$\alpha$ wavelength relative to the average efficiency of the
filter and about unity. Although this modification is not very
important, we re-calculate Ly$\alpha$ fluxes of the LAE candidates in
the SDF. When the estimated $f_\lambda^{\rm cont}$ is less than its
1-$\sigma$ uncertainty estimated from an error propagation in equation
(2), we set $f_\lambda^{\rm cont}=0$ in equation (1). Finally, we
converted $f_{\rm Ly\alpha}$ into Ly$\alpha$ luminosity which are
summarised in Table~3. In the table, we also list the continuum
luminosity density and the observed equivalent width of Ly$\alpha$ for
which we adopted the estimated 1-$\sigma$ uncertainty as an upper or
lower limit.

\begin{table}
 \caption[]{Estimated properties of $z\simeq6.5$ LAE candidates in the
 SSA22 field.}
 \setlength{\tabcolsep}{3pt}
 \footnotesize
 \begin{minipage}{\linewidth}
  \begin{tabular}{lccc}
   \hline
   Object & $L_{\rm Ly\alpha}$ $^a$ & $L_{\rm UV}$ $^b$ 
   & EW$_{\rm obs}$ $^c$ \\
   & ($10^{42}$ erg s$^{-1}$) & ($10^{28}$ erg s$^{-1}$ Hz$^{-1}$) 
   & ($10^2$ \AA) \\
   \hline
   SSA22-z6p5-1 & 5.74 & 8.5 & 2.4 \\
   SSA22-z6p5-2 & 5.91 & $<4.8$ & $>4.5$ \\
   SSA22-z6p5-3 & 5.26 & $<4.8$ & $>4.0$ \\
   SSA22-z6p5-4 & 4.25 & $<4.8$ & $>3.2$ \\
   SSA22-z6p5-5 & 4.20 & $<4.8$ & $>3.2$ \\
   SSA22-z6p5-6 & 4.15 & $<4.8$ & $>3.1$ \\
   SSA22-z6p5-7 & 3.99 & $<4.8$ & $>3.0$ \\
   SSA22-z6p5-8 & 3.09 & 4.8 & 2.3 \\
   SSA22-z6p5-9 & 3.89 & $<4.8$ & $>2.9$ \\
   SSA22-z6p5-10 & 3.89 & $<4.8$ & $>2.9$ \\
   SSA22-z6p5-11 & 3.85 & $<4.8$ & $>2.9$ \\
   SSA22-z6p5-12 & 3.66 & $<4.8$ & $>2.8$ \\
   \hline
   SSA22-z6p5-13 $^d$ & 8.16 & $<16$ & $>1.9$ \\
   SSA22-z6p5-14 $^d$ & 8.49 & $<12$ & $>2.6$ \\
   \hline
  \end{tabular}

  $^a$ Ly$\alpha$ luminosity estimated by equation (1).\\
  $^b$ UV continuum luminosity estimated by equation (2).\\
  $^c$ Observed equivalent width of Ly$\alpha$.\\
  $^d$ The derived UV luminosities are an upper limit because $z'$ flux
  densities are contaminated by a neighbouring object.
 \end{minipage}
\end{table}%

Figure~6 shows the cumulative LFs of Ly$\alpha$ of the LAE candidates in
the SSA22 field and in the SDF. There is a clear difference. As
summarised in Table~4, the number density of the $z\simeq6.5$ LAE
candidates with $L_{\rm Ly\alpha}>3\times10^{42}$ erg s$^{-1}$ in the
SSA22 is a factor of 2.5 smaller than that in the SDF. The number ratio
(SDF/SSA22) is 3.5 for $L_{\rm Ly\alpha}>5\times10^{42}$ erg s$^{-1}$ 
and 3.4 for $L_{\rm Ly\alpha}>7\times10^{42}$ erg s$^{-1}$. In
Figure~6, the SSA22 LF seems to be reproduced if the number density of
the SDF LF is reduced by a factor of 0.3. On the other hand, a factor
of 0.6 reduction of $L_{\rm Ly\alpha}$ in the SDF LF also provides an
agreement with the SSA22 LF. The reduction of $L_{\rm Ly\alpha}$
corresponds to a scenario with a different neutral fraction of hydrogen
in the two fields. We will discuss this point more in section 5.2.

\begin{figure}
 \begin{center}
  \includegraphics[width=7cm]{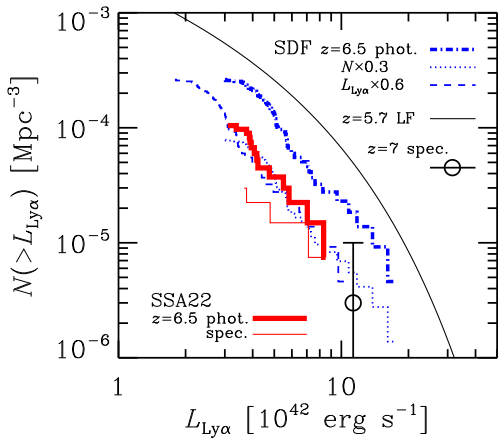}
 \end{center}
 \caption{Cumulative Ly$\alpha$ luminosity functions (LFs) of $z\ga6$
 LAEs. The thick solid and dot-dashed histograms are the LFs of
 photometric LAE candidates at $z=6.5$ in the SSA22 field and in the
 SDF, respectively. The thin solid histogram is the LF of spectroscopic
 LAE in the SSA22 field. No correction of the completeness has been
 applied. The dotted histogram is the SDF LF whose number density is
 reduced by a factor of 0.3. The dashed histogram is the SDF LF with a
 reduction factor of 0.6 for the Ly$\alpha$ luminosity. The thin solid
 curve is the LF for $z=5.7$ LAEs in the SDF by Shimasaku et al.~(2006).
 The open circle with an error-bar is the estimation based on one
 spectroscopic LAE at $z=7$ in the SDF by Ota et al.~(2008).}
\end{figure}

\begin{table}
 \caption[]{The observed number density in unit co-moving volume of
 $z\simeq6.5$ LAE candidates in SSA22 and SDF.}
 \setlength{\tabcolsep}{3pt}
 \footnotesize
 \begin{minipage}{\linewidth}
  \begin{tabular}{lcc}
   \hline
   $L_{\rm Ly\alpha}$ & $N$(SSA22)$^a$ & $N$(SDF)$^a$\\
   ($10^{42}$ erg s$^{-1}$) & ($10^{-5}$ Mpc$^{-3}$) 
       & ($10^{-5}$ Mpc$^{-3}$) \\
   \hline
   $>10$ & $0^{+1.4}$ & $2.3^{+1.6}_{-1.0}$ \\
   $>7$ & $1.5^{+2.0}_{-1.0}$ & $5.1^{+2.0}_{-1.5}$ \\
   $>5$ & $3.7^{+2.5}_{-1.6}$ & $12.9^{+2.9}_{-2.4}$ \\
   $>3$ & $10.4^{+3.6}_{-2.7}$ & $25.8^{+3.9}_{-3.4}$ \\
   \hline
  \end{tabular}

  $^a$ 1-$\sigma$ uncertainties are based on Poisson statistics (Gehrels
  1986).
 \end{minipage}
\end{table}%

\subsection{Number density of $i'$-drop galaxies}

Here we compare the number densities of $i'$-drop galaxies in the SSA22
and in the SDF. This comparison may be useful to resolve the reason why
the number of $z\simeq6.5$ LAE candidates in the SSA22 is so small
relative to that in the SDF. Namely, the small number of the LAE
candidates in the SSA22 may suggest that the field is just a void-like
region at $z\simeq6.5$. If so, the number density of $i'$-drop galaxies
may be also small. However, we should be care about the difference of
the depths along the line of sight for sampling $i'$-drop galaxies and
the LAE candidates: $\sim250$ $h_{0.7}^{-1}$ Mpc for $i'$-drop galaxies
but 42 $h_{0.7}^{-1}$ Mpc for the LAE candidates.

To select $i'$-drop galaxies among objects detected in the $z'$ band, we
adopt the following criteria for the SSA22 (SDF):
\begin{quote}
 \item[(a)] $25.50<z'<26.00$ ($5\sigma$), 
 \item[(b)] $i'-z'>1.5$ or $i'>27.53$ (27.84) ($2\sigma$), 
 \item[(c)] $B>27.89$ (28.38), $V>27.66$ (27.69), and $R>27.03$ (27.76) 
 ($3\sigma$), 
\end{quote}
where $\sigma$ indicates 1-$\sigma$ uncertainty of photometry in each
bands and the values in parentheses for the SDF. These criteria are
similar to those in \cite{nag04} and \cite{ota05}. We do not have
near-infrared data and cannot adopt any rest-frame UV colour criteria as
done by \cite{bou03} and \cite{shi05}. Thus, our sample of the $i'$-drop
objects is contaminated by Galactic M/L/T dwarfs. However, the
contamination is $<10\%$ for $z'>25.5$ AB \citep{ota05}. 
The survey area in the SSA22 is the same as the LAE candidates.

The resultant numbers of the $i'$-drop galaxies are 25 in the SSA22 and
36 in the SDF. The surface densities are $0.047\pm0.009$ arcmin$^{-2}$
in the SSA22 and $0.036\pm0.006$ arcmin$^{-2}$ in the SDF both which are
consistent with that in the SXDS field by \cite{ota05}. 
Therefore, the number density of $i'$-drop galaxies in the
SSA22 is `normal'. This may suggest that a large difference of the
number densities of the LAE candidates in the two fields is not caused
by different number densities of dark matter haloes. However, we should
caution ourselves that the sampling redshift of the $i'$-drop selection
is different from that of the LAE selection. On the other hand, the
spatial distribution of the $i'$-drop galaxies in the SSA22 seems to be
well correlated to that of the LAE candidates as shown in Figure~3. This
may indicate a similar redshift range of the $i'$-drop galaxies to the
LAE candidates, while we should wait for spectroscopic confirmations to
reach the final conclusion.

\subsection{Effect of smaller aperture size}

Because of finer FWHMs in our final images, the aperture sizes in our
photometry are smaller than those in the SDF. This difference may affect
the estimation of Ly$\alpha$ luminosity and the resultant LF unless
the Ly$\alpha$ emissions are point-like in our image. Indeed, some LAE
candidates seem to be spatially extended as found in Figure~4. To
examine this issue, we have performed the LAE selection again in the
images convolved with a Gaussian kernel so as that FWHMs of point-like
sources become $1.''0$ which is the same in the SDF images. The
selection criteria are the same as \S3, but the limiting magnitudes are
$\sim0.3$ mag shallower. The photometry was made within a diameter of
twice of the FWHM (i.e., $2.''0$). After removing objects affected by
spiders of bright stars, we have found 6 objects with NB912 $<25.76$
(5-$\sigma$) and 4 objects with $L_{\rm Ly\alpha}>5\times10^{42}$ erg
s$^{-1}$. This number is consistent with 3 objects found in the
finer FWHM images other than the objects SSA22-z6p5-13 and -14 whose
photometry affected by a foreground object. Therefore, the smaller
aperture sizes due to finer FWHMs do not affect our conclusion.

\section{Discussions}

\subsection{Cosmic variance}

We have found a large difference in the number densities of $z\simeq6.5$
LAE candidates in the SSA22 and in the SDF. Here we examine if this
difference can be explained by variance of the number density of dark
matter haloes, i.e. cosmic variance. For this aim, we assume the SDF LF
to be representative.

First, we calculate the expected number of $z\simeq6.5$ LAE candidates
in the SSA22 from the observed number in the SDF. As shown in the third
column of Table~5, the expected numbers, $N_{\rm exp}$, are a factor of
about 3 larger than the observed numbers, $N_{\rm obs}$. Next, we
estimate a fractional standard deviation by cosmic variance,
$\sigma_{\rm cv}$, according to \cite{som04}. As summarised in the
fourth column of Table~5, the $\sigma_{\rm cv}$ values become smaller
for less luminous LAE candidates because the bias parameters are
smaller. The survey volume of our SSA22 survey is smaller than that of
the SDF, and thus, the resultant $\sigma_{\rm cv}$ values are larger
than those estimated in \cite{kas06} for the SDF. Although we adopted
the lines for $z=6$ in figure 3 of \cite{som04} which are the closest to
$z=6.5$, the uncertainty caused by the redshift difference would be
small.

In order to examine the statistical significance of the smallness of the
observed numbers, we have performed a Monte-Carlo simulation as follows; 
Assuming the cosmic variance to be Gaussian, we first draw a random 
number from the Gaussian distribution with its mean of $N_{\rm exp}$ and
the standard deviation of $N_{\rm exp}\sigma_{\rm cv}$. Then, we draw
another random number from the Poisson distribution with its parameter
of the first random number and store the second random number as the
``expected observed number'', $N_{\rm obs}^{\rm exp}$. We repeat this
procedure 100,000 times for each Ly$\alpha$ luminosity bin. Finally, we
count the realizations whose $N_{\rm obs}^{\rm exp}$ is equal to or
smaller than the real observed number $N_{\rm obs}$. The resultant
probabilities are listed in the last column of Table~5. We find that the
probability to have $N_{\rm obs}$ of LAE candidates with 
$L_{\rm Ly\alpha}>5-3\times10^{42}$ erg s$^{-1}$ as small as in the
SSA22 field is less than 7\%, while the probability is higher for more
luminous LAEs due to a larger Poisson error. From the Monte-Carlo
simulation, we may conclude that the SSA22 field is a rare void at
$z\simeq6.5$.

However, we find no significant difference of the number density
of $i'$-drop galaxies in the SSA22 and the SDF. Indeed, the number of
$i'$-drop galaxies in the SSA22 expected from that in the SDF is
consistent as summarised in Table~6. Note that the $\sigma_{\rm cv}$ for
this case is smaller than those in the LAE case because the survey
volume is larger. On the other hand, the survey depth along the line of
sight for $i'$-drop galaxies is about 6 times larger than that for
LAEs. This point may explain the consistency of the number density of
$i'$-drop galaxies even if there is a void at $z\simeq6.5$ traced by
LAEs.

\begin{table}
 \caption[]{Number of $z\simeq6.5$ LAE candidates in the SSA22 survey
 expected from the SDF luminosity function and cosmic variance.}
 \setlength{\tabcolsep}{3pt}
 \footnotesize
 \begin{minipage}{\linewidth}
  \begin{tabular}{lcccc}
   \hline
   $L_{\rm Ly\alpha}$ & $N_{\rm obs}$ $^a$ & $N_{\rm exp}$ $^b$ 
   & $\sigma_{\rm cv}$ $^c$ & $P(\leq N_{\rm obs})$ $^d$\\
   ($10^{42}$ erg s$^{-1}$) & & & (\%) & (\%) \\
   \hline
   $>10$ & 0 & 3.1 & 49 & 12 \\
   $>7$ & 2 & 6.8 & 46 & 14 \\
   $>5$ & 5 & 17.4 & 41 & 6.5 \\
   $>3$ & 14 & 34.7 & 36 & 6.6 \\
   \hline
  \end{tabular}

  $^a$ Observed number of the LAE candidates in the SSA22 survey.\\
  $^b$ Number of the LAE candidates found in the SSA22 survey expected
  from the SDF result.\\
  $^c$ Fractional standard deviation by cosmic variance estimated by the
  method in \cite{som04} based on the observed number density in the SDF
  and the SSA22 survey volume.\\
  $^d$ Probability to have a number equal to or smaller than the
  observed number. This is estimated from the expected number by a
  Monte-Carlo simulation taking into account the Gaussian cosmic
  variance and Poisson error.
 \end{minipage}
\end{table}%

\begin{table}
 \caption[]{Same as Table~4 but for $i'$-drop galaxies.}
 \setlength{\tabcolsep}{3pt}
 \footnotesize
 \begin{minipage}{\linewidth}
  \begin{tabular}{lcccc}
   \hline
   $z'$ & $N_{\rm obs}$ & $N_{\rm exp}$ & $\sigma_{\rm cv}$
   & $P(\geq N_{\rm obs})$\\
   (AB) & & & (\%) & (\%) \\
   \hline
   25.5--26.0 & 25 & 19.5 & 27 & 23 \\
   \hline
  \end{tabular}
 \end{minipage}
\end{table}%

\subsection{Difference of neutral fraction and implications for reionization}

As an alternative scenario, if neutral hydrogen remains more in the
SSA22 than in the SDF at $z=6.5$, Ly$\alpha$ emission is more heavily
obscured in the SSA22, and thus, the number density of the observable
LAEs decreases in the SSA22. As found in Figure 6, the cumulative LF in
the SSA22 is almost reproduced if the Ly$\alpha$ luminosity in the SDF
is reduced by a factor of 0.6. According to \cite{san04}, the
Ly$\alpha$ transmission through the IGM is roughly proportional to
$1/x_{\rm HI}$, where $x_{\rm HI}$ is the hydrogen neutral fraction in
the IGM. Therefore, the factor of 0.6 reduction of Ly$\alpha$ luminosity
can be interpreted as a factor of about 2 larger $x_{\rm HI}$ in the
SSA22 than in the SDF.

\cite{kas06} have already found a significant reduction of the
Ly$\alpha$ luminosity at $z=6.5$ relative to $z=5.7$ in the SDF and
argue that the IGM is more neutral at $z=6.5$ than $z=5.7$. Their
estimation based on \cite{san04} model is $x_{\rm HI}<0.45$ at 
$z=6.5$.\footnote{\cite{kas06} obtained an upper limit on $x_{\rm HI}$.
The Ly$\alpha$ luminosity reduction (or IGM transmission of Ly$\alpha$)
could be accounted for by almost ionized IGM if there is no galactic
wind (see Fig.~25 in \citealt{san04}).} In the SSA22, there may be a
further factor of $\sim2$ reduction of Ly$\alpha$ luminosity. This is
translated into $x_{\rm HI}<0.9$ at $z=6.5$ in the SSA22. Therefore,
more than half hydrogen may be still neutral at $z=6.5$ in the SSA22
field.

The furthest known LAE is IOK-1 at $z=6.96$ found in the SDF by
\cite{iye06}. \cite{ota08} have constrained $x_{\rm HI}$ at $z=7$ in the
SDF based on this one object: $x_{\rm HI}=0.1$--0.6. The Ly$\alpha$
luminosity of this object is $1.1\times10^{43}$ erg s$^{-1}$. On the
other hand, we could not find any LAE candidates with 
$L_{\rm Ly\alpha}>1\times10^{43}$ erg s$^{-1}$ at $z=6.5$ in the SSA22.
The number density of $z=7$ LAE in the SDF is $\sim3\times10^{-6}$
Mpc$^{-3}$ for $L_{\rm Ly\alpha}>1\times10^{43}$ erg s$^{-1}$ which may
match with an extrapolation of the SSA22 LF shown in Figure~6. This may
imply that the SSA22 at $z=6.5$ is as neutral as the SDF at $z=7$.

If the $z=6.5$ IGM in the SSA22 field is more neutral than that in the
SDF and Ly$\alpha$ luminosity in the SSA22 field is reduced, an effect
may appear in the Ly$\alpha$ EW distribution. Figure~7 shows the
observed Ly$\alpha$ EW as a function of UV luminosity for $z=6.5$
photometric LAE samples in the SSA22 field and in the SDF. Although we
expect that the EWs of the SSA22 sample are systematically smaller than
those of the SDF sample, we can not find such a trend. On the other
hand, the UV luminosities in the SSA22 field may be systematically
smaller than those in the SDF, which may indicate that the small number
of the LAEs in the SSA22 field is due to a cosmic variance. However, in
any case, we can not extract any conclusion due to the small statistics
of the SSA22 sample. A future work with a much larger sample is required.

\begin{figure}
 \begin{center}
  \includegraphics[width=7cm]{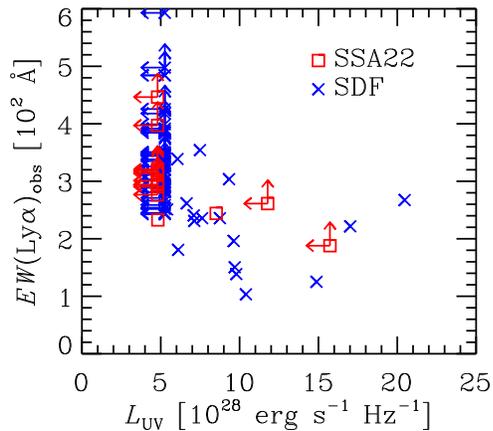}
 \end{center}
 \caption{Observed Ly$\alpha$ equivalent width as a function of UV
 luminosity. The squares are the $z=6.5$ LAE candidates in the SSA22
 field and the crosses are those in the SDF. The arrows indicate upper
 or lower limits for $L_{\rm UV}$ or $EW$, respectively.}
\end{figure}

\section{Conclusions}

This paper presents results of a deep survey of $z=6.5$ LAE candidates
in the SSA22 field based on the imaging data collected with
Subaru/S-Cam. We have selected the LAE candidates by the standard
technique based on a narrowband excess by an emission line and drop-outs
at shorter wavelengths. The adopted narrowband is NB912 which can
capture Ly$\alpha$ emission at $z=6.46$--6.57. We have also selected
$i'$-drop galaxies for a comparison. These galaxies exist at a similar
high redshift to but more extended range than the LAE candidates.

A summary of the observational findings is as follows;  
The number of the LAE candidates with $L_{\rm Ly\alpha}>3\times10^{42}$ 
erg s$^{-1}$ in the observed area of $\approx530$ arcmin$^2$ in the
SSA22 field is only 14. The co-moving number density of the LAE
candidates is a factor of 3 smaller than that in the SDF reported by
\cite{tan05} and \cite{kas06}. On the other hand, the number
densities of $i'$-drop galaxies in the two fields agree within the
Poisson error.

We have considered two possibilities accounting for the small number
density of the LAE candidates in the SSA22: (1) fluctuation due to the
large-scale structure (i.e. so-called cosmic variance) and (2)
fluctuation of the neutral fraction of hydrogen in the IGM. In the
cosmic variance scenario, the probability to have the small number
density as in the SSA22 field is 7\% if we assume the SDF to be
the cosmic average. Therefore, the SSA22 field at $z=6.5$ may be a rare
void. However, there is no evidence of such a void in the $i'$-drop
galaxies although their redshift range is different from the LAE
candidates. 

If the neutral fraction in the IGM is higher in the SSA22 than in the
SDF, a part of Ly$\alpha$ emission is more heavily obscured in the
SSA22. Indeed, a factor of about 2 reduction of Ly$\alpha$ luminosity in
the SDF provides a good fit of the SSA22 Ly$\alpha$ LF. This is
interpreted as a factor of about 2 higher neutral fraction in the SSA22
field than in the SDF. Even in the SDF, $z=6.5$ Ly$\alpha$ LF indicates
a factor of 2 reduction of Ly$\alpha$ luminosity relative to $z=5.7$ and
this reduction puts a constraint on the neutral fraction at $z=6.5$ as
$x_{\rm HI}<0.45$ \citep{kas06}. Therefore, we may obtain 
$x_{\rm HI}<0.9$ at $z=6.5$ in the SSA22. This is similar to that at
$z=7$ in the SDF reported by \cite{ota08}. 

In any case, the smallness of the number density of the $z=6.5$ LAE
candidates in the SSA22 clearly shows that there is a large fluctuation
of the LAE number density at this high redshift as already suggested by
\cite{hu04,hu05,hu06,hu10}. \cite{ouc08} also reports a factor of 5
variation in the number densities of $z=5.7$ LAEs among 5 fields-of-view
of the S-Cam. Therefore, we need a much wider survey area than the
surveys reported here and previously in order to obtain a robust mean LF
of $z=6.5$ LAEs and a robust mean neutral fraction in the IGM at the
redshift.

\section*{Acknowledgments}

We greatly thank Nobunari Kashikawa for valuable discussions.
A.K.I. is supported by the Institute for Industrial Research, Osaka
Sangyo University and by KAKENHI (the Grant-in-Aid for Young Scientists
B: 19740108) by The Ministry of Education, Culture, Sports, Science and
Technology (MEXT) of Japan.

\label{lastpage}

\end{document}